\documentclass{aa}
\usepackage{amsmath,array}

\usepackage{txfonts}
\usepackage{sansmath}

\begin{document}

\title{The Galactic Center Molecular Cloud Survey}
\subtitle{II. A Lack of Dense Gas \& Cloud Evolution along Galactic
  Center Orbits}

\author{Jens~Kauffmann \inst{1}
  \and 
  Thushara~Pillai \inst{1}
  \and
  Qizhou~Zhang \inst{2}
  \and
  Karl~M.~Menten \inst{1}
  \and
  Paul~F.~Goldsmith \inst{3}
  \and
  Xing~Lu \inst{2,4}
  \and
  Andr\'es~E.~Guzm\'an \inst{5}
  \and
  Anika~Schmiedeke \inst{6}
}

\institute{Max--Planck--Institut f\"ur Radioastronomie, Auf dem
  H\"ugel 69, 53121 Bonn, Germany
  \and
  Harvard--Smithsonian Center for Astrophysics, 60 Garden Street,
  Cambridge, MA 02138, USA
  \and
  Jet Propulsion Laboratory, California Institute of Technology, 4800
  Oak Grove Drive, Pasadena, CA 91109, USA
  \and
  School of Astronomy and Space Science, Nanjing University, 22 Hankou
  Road, Nanjing 210093, China
  \and
  Departamento de Astronom\'ia, Universidad de Chile, Camino el
  Observatorio 1515, Las Condes, Santiago, Chile
  \and
  I.\ Physikalisches Institut, Universit\"at zu K\"oln, Z\"ulpicher
  Strasse 77, 50937 K\"oln, Germany}

\abstract{We present the first systematic study of the density
  structure of clouds found in a complete sample covering all major
  molecular clouds in the Central Molecular Zone (CMZ; inner
  $\sim{}200~\rm{}pc$) of the Milky Way. This is made possible by
  using data from the Galactic Center Molecular Cloud Survey (GCMS),
  the first study resolving all major molecular clouds in the CMZ at
  interferometer angular resolution. We find that many CMZ molecular
  clouds have unusually shallow density gradients compared to regions
  elsewhere in the Milky Way. This is possibly a consequence of weak
  gravitational binding of the clouds. The resulting relative absence
  of dense gas on spatial scales $\sim{}0.1~\rm{}pc$ is probably one
  of the reasons why star formation (SF) in dense gas of the CMZ is
  suppressed by a factor $\sim{}10$, compared to solar neighborhood
  clouds. Another factor suppressing star formation are the high SF
  density thresholds that likely result from the observed gas
  kinematics. Further, it is possible but not certain that the star
  formation activity and the cloud density structure evolve
  systematically as clouds orbit the CMZ.}


\maketitle

%\tableofcontents

\defcitealias{kauffmann2016:gcms_i}{Paper~I}

\section{Introduction\label{sec:introduction}}
The Central Molecular Zone (CMZ) --- i.e.\ the inner
$\sim{}200~\rm{}pc$ of our Galaxy --- is a star--forming environment
with very extreme physical properties. It contains molecular clouds
that have unusually high average densities
$\gtrsim{}10^4~\rm{}cm^{-3}$ on spatial scales of about 1~pc
(Sec.~\ref{sec:densities}), are subject to a high confining pressure
of $10^{6~{\rm{}to}~7}~\rm{}K\,cm^{-3}$
\citep{yamauchi1990:cmz-plasma, spergel1992:pressure-bulge,
  muno2004:diffuse-xray}, and are penetrated by strong magnetic fields
with strengths $\sim{}5~\rm{}mG$
\citep{pillai2015:magnetic-fields}. See
\citeauthor{kauffmann2016:gcms_i} (\citeyear{kauffmann2016:gcms_i};
hereafter \citetalias{kauffmann2016:gcms_i} of this series) for a brief
recent summary of CMZ physical conditions.

Research into CMZ star formation is critical because (\textit{i})~this
permits us to test models of star formation (SF) physics in extreme
locations of the SF parameter space, and (\textit{ii})~it potentially
provides us with well--resolved nearby templates that help to decode
the processes acting in starburst galaxies in the nearby and distant
universe. For these reasons we launched the Galactic Center Molecular
Cloud Survey (GCMS), the first systematic study resolving all major
CMZ molecular clouds at interferometer angular resolution. The GCMS
data are published in a series of
studies. \citetalias{kauffmann2016:gcms_i} centers on the study of
cloud kinematics and investigations into the SF activity in specific
CMZ clouds. In this second paper we focus on studies of the density
structure of the clouds observed.

Several peculiar factors influence the density structure of CMZ
clouds. One interesting aspect is that clouds at galactocentric radii
of about $20~{\rm{}to}~200~\rm{}pc$ are generally subject to
\emph{compressive tidal forces in the radial direction}\footnote{The
  classical reviews by, e.g., \citet{gusten1989:review} and
  \citet{morris1996:cmz-review} state that clouds are subject to
  disruptive shear. This argument is based on an analysis by
  \citet{gusten1980:h2co-cmz}, who adopted a CMZ mass profile
  $m\propto{}r^{1.2}$ based on the best data available then. Today's
  research suggests a much steeper relation $m\propto{}r^{2.2}$,
  though. Tidal forces are compressive in this situation. See Sec.~6
  of \citet{lucas2015:thesis} for a detailed discussion of tidal
  forces in the CMZ, and \citet{renaud2009:compressive-tides} for a
  general analysis of the tidal tensor.} (e.g., Fig.~6.2 of
\citealt{lucas2015:thesis}) because the gravitational force
$F_{\rm{}g}\propto{}m/r^2$ \emph{increases} with increasing
galactocentric radius for the observed CMZ mass profile, where
$m\propto{}r^{2.2}$ (\citealt{launhardt2002:cmz-potential}; see
\citealt{kruijssen2014:orbit} for the measurement of the power--law
slope in the radial interval of about 10~to~100~pc that is of interest
here). The clouds are further compressed by the high external
pressure: gas temperatures of typically $50~{\rm{}to}~100~\rm{}K$ in
CMZ clouds (\citealt{guesten1981:nh3-cmz, huettemeister1993:nh3-cmz,
  ao2013:cmz-temperatures, mills2013:widespread-hot-nh3,
  ott2014:cmz-atca, ginsburg2015:cmz-gas-temperatures}; also see
\citealt{riquelme2010:isotopes, riquelme2012:temp-loop-interceptions})
imply $\rm{}H_2$ densities $\gtrsim{}10^4~\rm{}cm^{-3}$ to
balance\footnote{We thank William Lucas (School of Physics and
  Astronomy, University of St.~Andrews) for pointing this out.} the
aforementioned pressure of
$10^{6~{\rm{}to}~7}~\rm{}K\,cm^{-3}$. Finally, strong and widespread
SiO emission tracing shocks \citep{martin-pintado1997:sio-cmz,
  huettemeister1998:shocks, riquelme2010:survey}, prevalence of
molecules likely ejected from grain surfaces via shocks
\citep{requena-torres2006:corganic-mols,
  requena-torres2008:coxygen-coms}, and collisionally--excited
methanol masers (\citealt{mills2015:cmz-masers}; also see
\citealt{menten2009:g1.6}, though) suggest that much of the gas in the
CMZ is subject to violent gas motions, such as cloud--cloud collisions
at high velocities. These motions might further compress the clouds.

This combination of tidal action, external pressure, and cloud
interactions probably results in the high average gas densities
observed in the CMZ. What is so far not known is the density structure
of CMZ clouds on spatial scales of about 0.1 to 1~pc. The
characterization of cloud structure on these small spatial scales is
one of the central goals of the GCMS.

Detailed knowledge of the structure of CMZ dense gas on small
spatial scales is crucial for our understanding of key properties of
the CMZ. In particular, it is generally established that, relative to
the solar neighborhood \citep{heiderman2010:sf-law,
  lada2010:sf-efficiency, evans2014:sfr-nearby-clouds}, star formation
in the dense gas of the CMZ is suppressed by an order of magnitude
(\citealt{guesten1983:h2o-masers, caswell1983:water-masers,
  caswell1996:masers-methanol, taylor1993:cmz-water-masers,
  lis1994:m0.25, lis2001:ir-spectra, lis1998:m025,
  immer2012:multi-wavelength-cmz, immer2012:recent-sfr,
  longmore2012:sfr-cmz, kauffmann2013:g0.253}; \citetalias{kauffmann2016:gcms_i}). It has been
argued that the aforementioned star formation relations describe both
the solar neighborhood and the integral star formation activity of
entire galaxies \citep{gao2004:hcn, lada2012:sf-laws}.

\begin{figure}
\includegraphics[width=\linewidth]{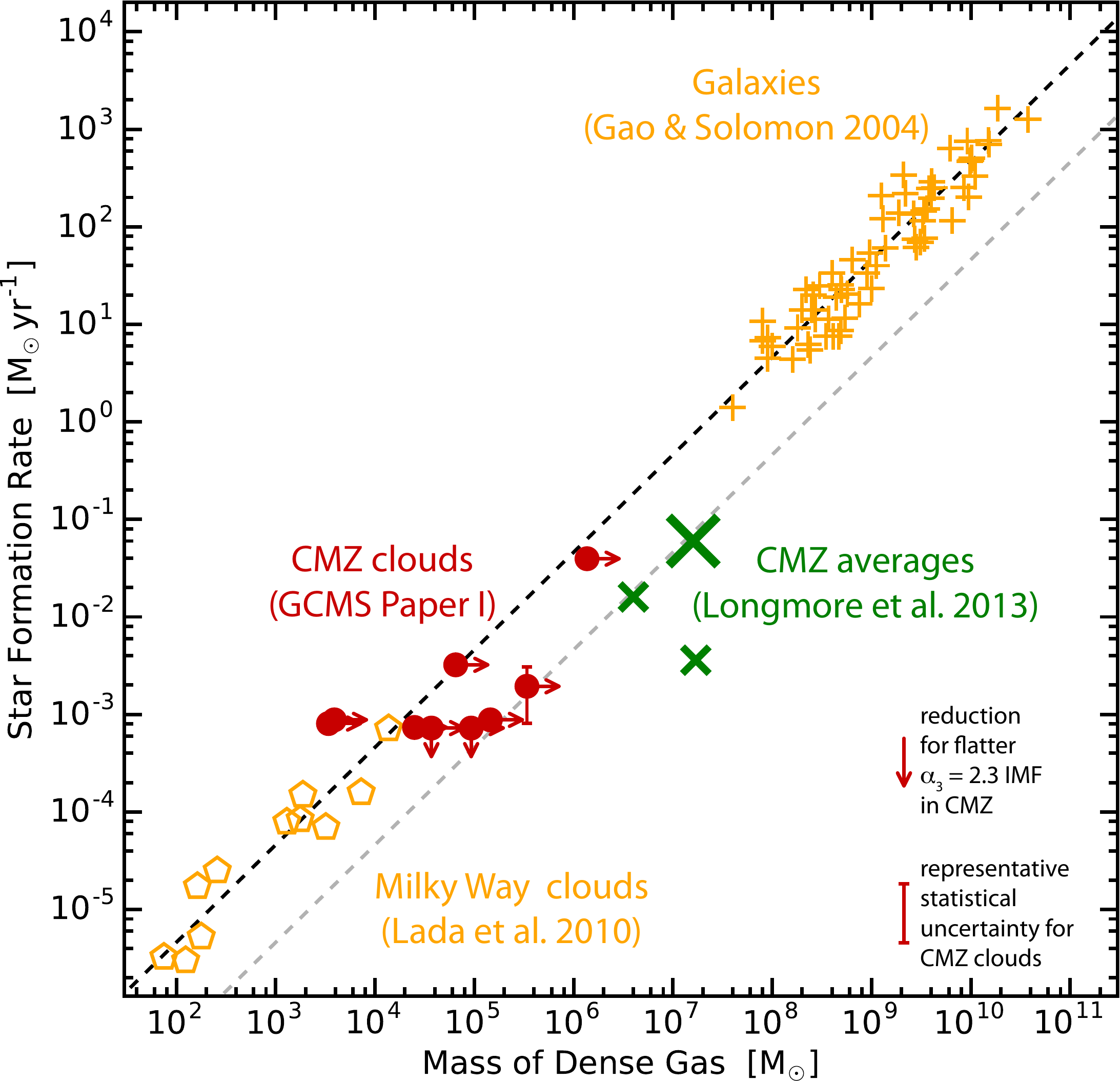}
\caption{Observed star formation rates vs.\ the mass of dense gas
  residing at visual extinctions $A_V>7~\rm{}mag$. \emph{Yellow
    diamonds} give the properties of molecular clouds within about
  500~pc from sun compiled by \citet{lada2010:sf-efficiency}. The
  \emph{yellow crosses} present the \citet{gao2004:hcn} data for star
  formation in entire galaxies, as re--calibrated by
  \citet{lada2012:sf-laws}. The \emph{large green cross} gives the CMZ
  star formation rate for the $|\ell{}|\le{}1\degr$ region largely
  explored by this paper, while the \emph{smaller green crosses} hold
  for further regions explored by \citet{longmore2012:sfr-cmz}. The
  \emph{red bullets} give masses and star formation rates for
  individual CMZ clouds discussed in this paper. See \citetalias{kauffmann2016:gcms_i} for
  details and the uncertainties illustrated by \emph{red error bars and
    arrows} shown in the lower right corner. The \emph{black dashed
    line} indicates a fit to the \citet{lada2010:sf-efficiency} data
  taken from the same publication. The \emph{gray dashed line} gives a
  relation with a star formation rate lower by a factor
  10.\label{fig:sf-mass}}
\end{figure}

The CMZ thus provides a unique and important laboratory to study
suppressed SF in dense gas (Fig.~\ref{fig:sf-mass}). This is an
important endeavor: similar suppression mechanisms might well affect
the growth of galaxies elsewhere in the cosmos. Several theoretical
research projects have been launched to understand the observed trends
in star formation. \citet{kruijssen2013:sf-suppression-cmz} review
analytical models of suppressed SF, while
\citet{bertram2015:sf-efficiencies} conduct numerical studies of
clouds subjected to CMZ conditions. \citet{krumholz2015:alpha-disk}
propose that inward transport of gas in a viscous disk could explain
many of the cloud properties observed at $|\ell|\lesssim{}3\degr$.

There are essentially three ways to inhibit star formation: by
suppressing the formation of dense molecular clouds, by suppressing
the formation of cores of $\sim{}0.1~\rm{}pc$ size that could
efficiently form individual stars (or small groups), or by suppressing
the collapse of these cores into
stars. \citetalias{kauffmann2016:gcms_i} establishes that a large
number of massive and dense clouds exist, and that SF is suppressed
\emph{inside} these clouds. Thus we can reject the option of pure
suppressed cloud formation. Further, in \citet{kauffmann2013:g0.253}
we use the first resolved maps of dust emission of the G0.253+0.016
cloud (a.k.a.\ the ``Brick'') to demonstrate for the first time that
at least one CMZ cloud is essentially devoid of significant dense
cores that could efficiently form a large number of stars. This
indicates that the suppression of dense core formation suppresses SF
activity. This picture is confirmed by subsequent studies of
G0.253+0.016 that also reveal little dense gas in this cloud
\citep{johnston2014:g0.253, rathborne2014:g0253-pdf,
  rathborne2015:g0253-alma} and characterize this aspect of cloud
structure via probability density functions (PDFs) of gas column
density. These studies do not, however, explore whether the density
structure of G0.253+0.016 is representative of the average conditions
in CMZ molecular clouds. Such research is the goal of the present
paper. In this study we use mass--size relations for molecular clouds
to characterize their density structure: relations
$m\propto{}r_{\rm{}eff}^b$ between the masses and effective radii of
cloud fragments do, for example, imply density gradients
$\varrho{}\propto{}r^{b-3}$ under the assumption of spherical symmetry
(see Sec.~\ref{sec:density-gradients} and
\citealt{kauffmann2010:mass-size-ii}). This analysis reveals that
unusually steep mass--size relations prevail in the CMZ, indicating
unusually shallow density gradients within clouds.

Here we use dust emission data from the Submillimeter Array (SMA; near
280~GHz frequency) and the Herschel Space Telescope (at 250 to
$500~\rm{}\mu{}m$ wavelength) for a first comprehensive survey of the
density structure of several CMZ clouds. The data are taken from
\citetalias{kauffmann2016:gcms_i}. Two conclusions from that study
are of particular importance for the current research.
\begin{itemize}
\item It has been established for many years that CMZ molecular clouds
  have unusually large velocity dispersions on spatial scales
  $\gtrsim{}1~\rm{}pc$, when compared to clouds elsewhere in the Milky
  Way. \citetalias{kauffmann2016:gcms_i} demonstrates that the velocity dispersion on smaller
  spatial scales becomes similar for clouds inside and outside of the
  CMZ. In other words, random ``turbulent'' gas motions in the dense
  gas of CMZ clouds are relatively slow.
\item Previous work shows that the star formation in the dense gas
  residing in the CMZ is suppressed by a factor $\sim{}10$ when
  compared to dense gas in the solar neighborhood. In \citetalias{kauffmann2016:gcms_i} we show
  that this suppression also occurs \emph{within} dense and
  well--defined CMZ molecular clouds.
\end{itemize}
In the present paper the new information on cloud density structure
derived below is combined with these previous results on cloud
kinematics and the star formation activity.  Given our comprehensive
sample, the GCMS now allows for the first time to explore how cloud
properties vary within the CMZ. In particular, this permits us to test
the scenario for cloud evolution proposed by
\citet{kruijssen2013:sf-suppression-cmz} and
\citet{longmore2013:cmz-cluster-progenitors}. This picture of cloud
evolution builds on the idea that all major CMZ clouds move along one
common orbit that might be closed \citep{molinari2011:cmz-ring} or
consist of open eccentric streams \citep{kruijssen2014:orbit}. It is
then plausible to think that certain positions along this CMZ orbit
are associated with particular stages in the evolution of clouds. Here
we can test this picture.\medskip

\noindent{}The paper is organized as
follows. Section~\ref{sec:observations-processing} describes our
observational data and their processing. In Sec.~\ref{sec:densities}
we show that all major CMZ clouds have high average densities on
spatial scales $\gtrsim{}1~\rm{}pc$. However, many of the clouds have
unusually shallow density gradients on spatial scales
$\lesssim{}1~\rm{}pc$ (Sec.~\ref{sec:density-gradients}), and most
clouds contain relatively little gas on small spatial scales
$\lesssim{}0.1~\rm{}pc$ where individual stars and small groups can
form efficiently (Sec.~\ref{sec:dense-gas-fraction}). Several of the
CMZ clouds appear to be only marginally bound by self--gravity
(Sec.~\ref{sec:virial-parameter}). The structures on spatial scales
$\lesssim{}1~\rm{}pc$, however, appear to have a high chance to be
bound. Section~\ref{sec:cloud-evolution} combines these results to
examine whether clear evolutionary trends exist among CMZ clouds. We
present a summary in Sec.~\ref{sec:summary}.

\section{Observations \& Data
  Processing\label{sec:observations-processing}}
Our sample selection is described in \citetalias{kauffmann2016:gcms_i}. We essentially image all
major CMZ clouds with masses exceeding about
$3\times{}10^4\,M_{\sun}$. Specifically this includes the Sgr~C,
$20~\rm{}km\,s^{-1}$, $50~\rm{}km\,s^{-1}$, G0.253+0.016, Sgr~B1--off,
and Sgr~D clouds. We here also include data on the Sgr~B2 cloud (that
is missing from our original sample due to its large size) taken from
\citet{schmiedeke2016:sgrB2-dust}. Additional ancillary archival
information is collected for the Dust Ridge C and D clouds.

\citetalias{kauffmann2016:gcms_i} explains that Sgr~D is likely a foreground or background
object that is not physically related to the CMZ. We therefore handle
the information on Sgr~D with care. In particular we only present data
on this region if it can be clearly singled out using labels. Data on
Sgr~D are ignored otherwise.

\citetalias{kauffmann2016:gcms_i} describes the calibration and imaging of dust and
$\rm{}N_2H^+$ (3--2) line emission data from the Submillimeter Array
(SMA) and the Atacama Pathfinder Experiment (APEX). Interferometer
data are imaged jointly with single--dish observations in order to
sample all spatial scales present in our targets. In part this uses
data from the ATLASGAL survey using the LABOCA instrument on APEX
\citep{schuller2009:atlasgal}.  \citetalias{kauffmann2016:gcms_i} also details the calibration
and imaging of dust continuum emission data from the Herschel Space
Telescope. Here we briefly describe how the structure apparent in
these maps is characterized.

\subsection{Characterization of Dust Continuum
  Emission\label{sec:results-continuum}}
\begin{figure*}
\centerline{\includegraphics[width=\linewidth]{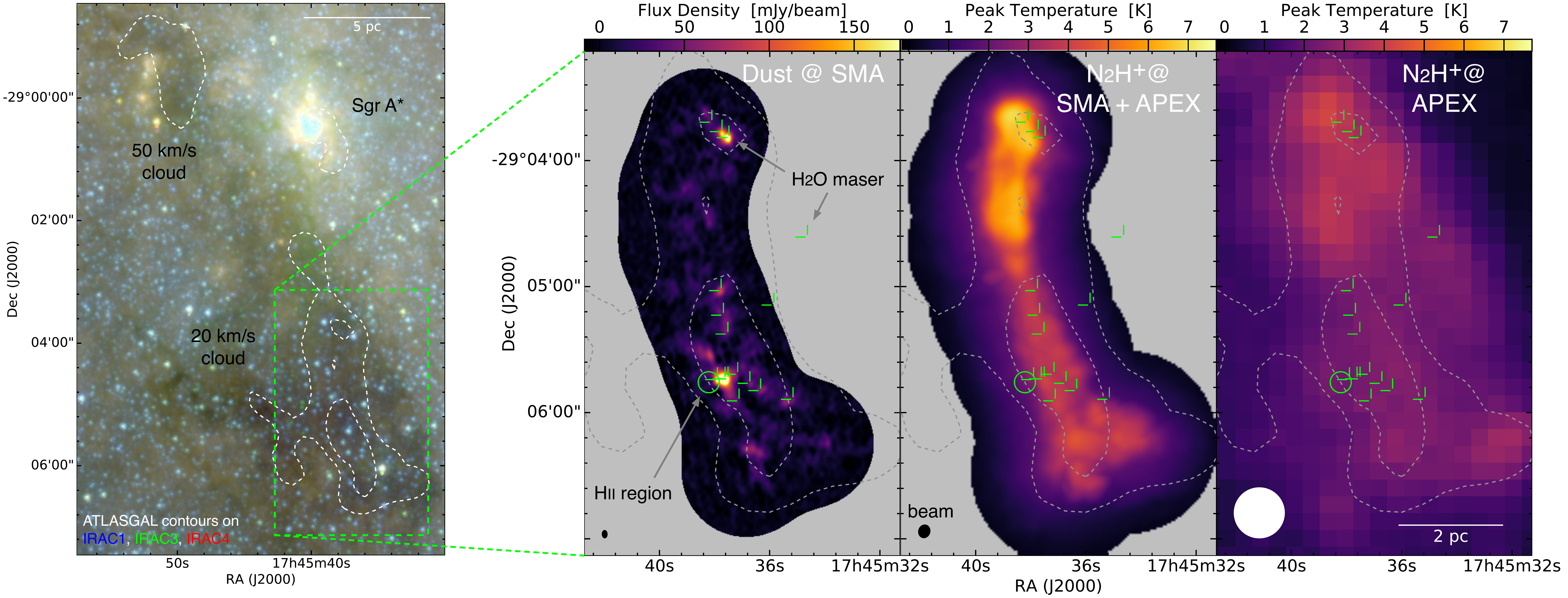}}
\caption{Example observations of the $\rm{}20~\rm{}km\,s^{-1}$
  cloud. The \emph{left panel} shows Spitzer IRAC data. The
  \emph{right panel} presents dust continuum emission near
  $280~\rm{}GHz$ as observed by the SMA, combined APEX and SMA
  observations of the $\rm{}H_2H^+$~(3--2) transition, and APEX--only
  observations of the $\rm{}H_2H^+$~(3--2) transition covering a
  larger area. All panels are overlaid with contours of single--dish
  dust continuum emission at $870~\rm{}\mu{}m$ wavelength from APEX at
  arbitrary chosen levels. Note that we show SMA data \emph{not}
  corrected for primary beam and missing extended emission to improve
  our visualization. The \emph{green circle} indicates the location of
  an \ion{H}{ii}~region found by \citet{ho1985:cmz_radio}, while
  \emph{green crosses} give the location of water masers discovered by
  \citet{lu2015:20kms}.\label{fig:20kms}}
\end{figure*}

Figure~\ref{fig:20kms} presents one of the SMA maps of dust emission
from \citetalias{kauffmann2016:gcms_i}. It serves as an example for the data we exploit in this
paper.  Here we use the data together with zero--spacing information
from single--dish APEX observations folded in. Noise levels and beam
shapes (the size is of order $2\arcsec$) are summarized in
Table~3 of \citetalias{kauffmann2016:gcms_i}.  We use the formalism from Appendix~A in
\citet{kauffmann2008:mambo-spitzer} to convert the dust emission
observations to estimates of masses and column densities. Dust
temperatures of $20~\rm{}K$ are adopted, following Herschel--based
estimates $\sim{}20~\rm{}K$ derived in
\citetalias{kauffmann2016:gcms_i}. \citet{ossenkopf1994:opacities} dust opacities for thin ice
mantles that have coagulated at a density of $10^6~\rm{}cm^{-3}$ for
$10^5~\rm{}yr$, are approximated as
\begin{equation}
  \kappa_{\nu} = 0.016~{\rm{}cm^2\,g^{-1}} \cdot{}
                            (\lambda/\text{mm})^{-1.75}
\end{equation}
following \citet{battersby2011:cluster-precursors}. \emph{We decrease
  these opacities by a factor 1.5} to be consistent with previous
dust--based mass estimates (see
\citealt{kauffmann2010:mass-size-i}). The resulting sensitivity to
mass and $\rm{}H_2$ column density depends on the beam size and noise
level (see Appendix~A in \citealt{kauffmann2008:mambo-spitzer}). For a
representative noise level of $4~\rm{}mJy$ per
$3\farcs{}3\times{}2\farcs{}3$ beam the $5\sigma$--noise--level
corresponds to mass and $\rm{}H_2$ column density sensitivities of
$21\,M_{\sun}$ and $7\times{}10^{22}~\rm{}cm^{-2}$, respectively. A
representative beam of $3\arcsec$ diameter corresponds to a linear
scale of $0.12~\rm{}pc$.

We caution that the dust properties in the CMZ might differ from those
prevalent in the solar neighborhood. For example, the relative Fe
abundance in CMZ stars might exceed the solar neighborhood value by a
factor $10^{0.11\pm0.15}\approx{}1.3$
\citep{ryde2014:cmz-abundances}. This suggests that the dust opacity
might be larger by a similar factor. However, the metallicity of CMZ
stars is highly diverse, including sub--solar values
\citep{ryde2014:cmz-abundances, do2015:cmz-abundances}. In this
situation we choose to adopt the aforementioned dust opacities that
are representative for the solar neighborhood. A possible increase in
dust opacities by a factor two would result in decreases of
dust--derived masses by the same factor.

The interferometer maps are largely devoid of significant
emission. This is a main feature of G0.253+0.016 that is already
reported by \citet{kauffmann2013:g0.253}. The new observations now
show that this relative absence of bright continuum emission is a
general feature of CMZ molecular clouds. A more quantitative
discussion of this trend is provided in
Sec.~\ref{sec:density-gradients}.

We characterize the dust emission using dendrograms
\citep{rosolowsky2008:dendrograms}. In practice we run the ASTRODENDRO
package\footnote{\url{http://www.dendrograms.org}} on the intensity
map resulting from the combination of SMA and APEX data. Intensities
and flux densities are converted into column densities and masses as
explained above. In essence this processing determines the mass and
size for every closed contour in the continuum map (e.g., Fig.~1 of
\citealt{kauffmann2010:mass-size-i}). Here we refer to these
structures as ``fragments'', independent of their size. The fragment
areas $A$ are converted into effective radii following
$r_{\rm{}eff}=(A/\pi)^{1/2}$. We follow the emission down to the
contour exceeding the noise level by a factor 3. Local maxima are
assumed to be significant if the depth of the saddle point separating
them from other local maxima exceeds the noise level by a factor
3. Only contours containing at least 10 pixels are considered,
where pixels have a size of $0\farcs{}4$. We
ignore unresolved structures, i.e., fragments with $r_{\rm{}eff}$
smaller than twice the beam radius. Such structures of small size form
parts of larger fragments that are extracted in our search.\medskip

\noindent{}We complement this information on the clouds with
information from the Herschel--based column density maps. Section~2.4
of \citetalias{kauffmann2016:gcms_i} describes how we determine the outer radius and total mass
of every cloud, as well as the peak mass per beam of $37\arcsec$,
corresponding to $r_{\rm{}eff}=0.74~\rm{}pc$. Table~1 of \citetalias{kauffmann2016:gcms_i}
presents these measurements.

For reference we illustrate the conditions in Orion using a
Herschel--based column density map of that region that is processed in
the same way as the CMZ data. See \citetalias{kauffmann2016:gcms_i} for details. We use those
data to characterize Orion on spatial scales
$\gtrsim{}0.5~\rm{}pc$. The Bolocam data from
\citet{kauffmann2013:g0.253} are used to assess the mass reservoir on
smaller spatial scales in this region. Dendrograms are used to
characterize these maps. We adopt a dust temperature of 25~K for the
analysis of the Bolocam maps of the Orion~KL region on spatial scales
$\lesssim{}0.1~\rm{}pc$ \citep{lombardi2014:herschel-planck-orion},
consistent with temperatures of 25~to~30~K our maps show for dense gas
immediately north of Orion~KL.\medskip

\noindent{}Figure~\ref{fig:mass-size-analysis}(a) presents the results
from this analysis. The clouds have total integrated masses between
$2.5\times{}10^4\,M_{\sun}$ and $3.4\times{}10^5\,M_{\sun}$ --- when
excluding Sgr~D, for which no total mass could be determined due to
insufficient contrast between cloud and background (see \citetalias{kauffmann2016:gcms_i}). These
mass reservoirs are enclosed in radii of 1~to~7~pc. It is already
obvious that on large spatial scale fragments within these clouds are
very massive --- and therefore dense --- compared to fragments of
similar size in Orion. A detailed discussion is presented in
Secs.~\ref{sec:densities} and \ref{sec:dense-gas}.

\subsection{Characterization of \sansmath{}$N_2H^+$ Line
  Emission\label{sec:line-emission}}
Here we briefly summarize procedures that are more completely
described in Sec.~3.3 of \citetalias{kauffmann2016:gcms_i}. We think that the hyperfine
structure of the $\rm{}N_2H^+$ (3--2) line is unlikely to have a
significant impact on the observed line structure. A compact summary
is, for example, provided by Table~A1 of
\citet{caselli2002:l1544_i}. This demonstrates that hyperfine
satellites with velocity offsets exceeding $\pm{}0.6~\rm{}km\,s^{-1}$
only contain a few percent of the integrated relative
intensities. Satellites within $\pm{}0.6~\rm{}km\,s^{-1}$ offset from
the line reference frequency might broaden the observed lines by about
$1~\rm{}km\,s^{-1}$, but the hyperfine structure is unlikely to have
further impact on the observed spectra.

We characterize the cloud kinematics using several methods. First, we
derive the velocity dispersion for the entire clouds. For this we
average all single--dish APEX spectra available for a given cloud. The
velocity dispersion is calculated as the second moment of these
spectra and reported in Table~4 of \citetalias{kauffmann2016:gcms_i}. Second, we segment the
emission in the maps combining single--dish and interferometer data on
spatial scales $\sim{}1~\rm{}pc$ by drawing an iso--intensity surface
in position--position--velocity ($p$--$p$--$v$) space at a threshold
intensity corresponding to $1/3$ of the peak intensity of the
respective map. Contiguous regions within these surfaces are
considered to be coherent structures. We refer to these structures as
``clumps''. The size of these clumps is characterized via the
effective radius $r_{\rm{}eff}=(A/\pi)^{1/2}$, where $A$ is the area
on the sky that contains all volume elements belonging to an extracted
structure. Spectra integrated within the surfaces are used to obtain
velocity dispersions. The latter property is only calculated if the
peak intensity of a clump exceeds the threshold intensity for clump
selection by a factor 2. Results are shown in Fig.~6 of
\citetalias{kauffmann2016:gcms_i}. Third, we find all significant local maxima in $p$--$p$--$v$
space that are by a factor 5 above the noise, and that are separated
from other significant maxima by troughs with a minimum depth
$\ge{}1~\rm{}K$. This effectively selects structures at the spatial
scale of the beam, i.e.\ $\sim{}0.1~\rm{}pc$. The spectra towards
these locations are fit by multi--component Gaussian curves in order
to obtain the velocity dispersions for the most narrow lines present
in the clouds. This yields velocity dispersions of
$0.6~{\rm{}to}~2.2~\rm{}km\,s^{-1}$.

\section{All CMZ Clouds have high Average
  Densities\label{sec:densities}}
\begin{figure}
\includegraphics[width=\linewidth]{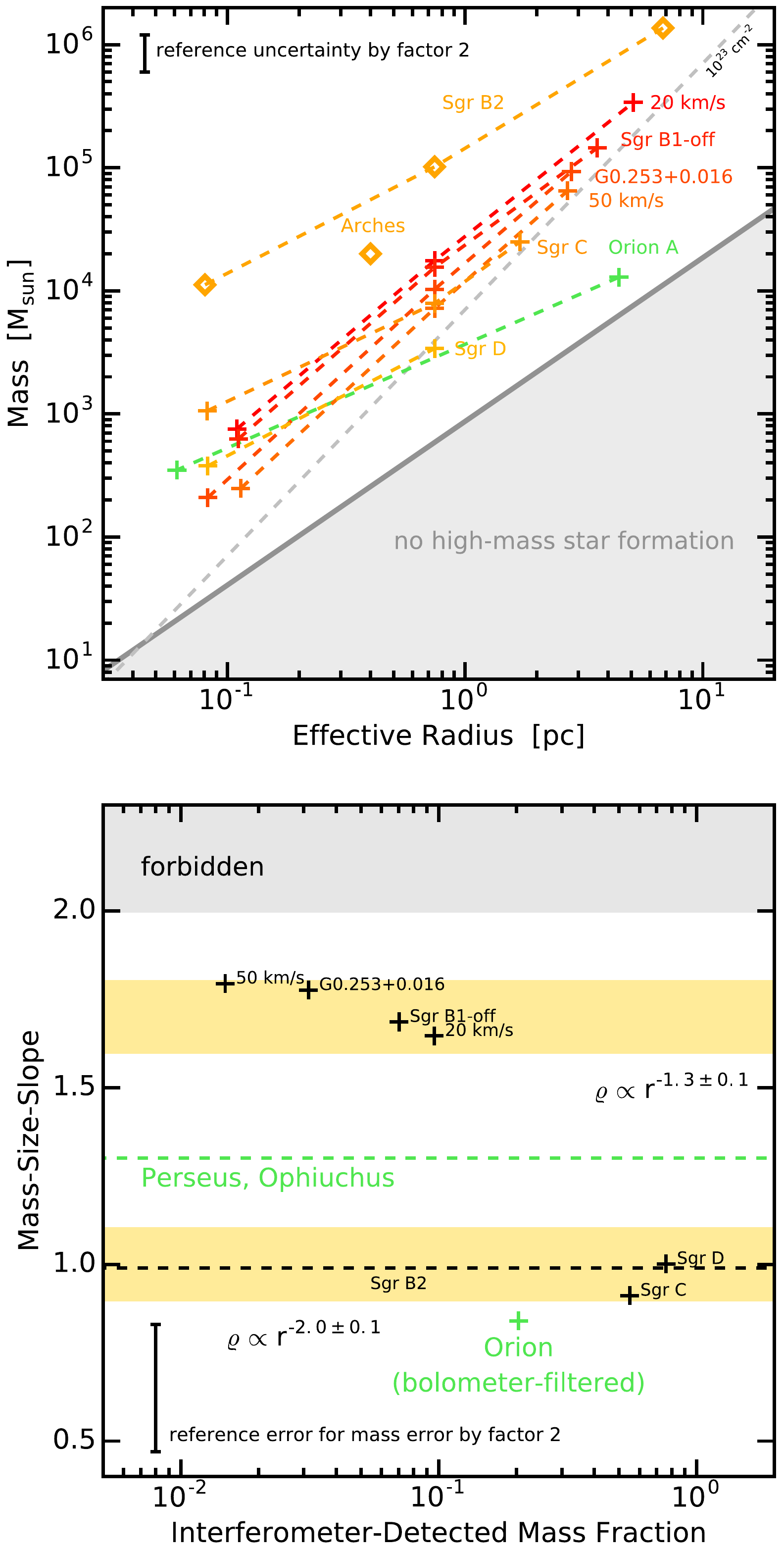}
\caption{Analysis of the density structure of CMZ clouds. The
  \emph{top panel} presents the mass--size data extracted for
  apertures of various sizes. See Sec.~\ref{sec:results-continuum} for
  details. For reference, \emph{gray shading} indicates the region of
  the parameter space where high--mass stars cannot form
  \citep{jenthu2010:irdcs}. The \emph{gray dashed line} indicates a
  mean $\rm{}H_2$ column density of $10^{23}~\rm{}cm^{-2}$. The
  \emph{ordinate} in the \emph{lower panel} gives the mass--size slope
  ${\rm{}d}\,\ln(m)/{\rm{}d}\,\ln(r)$ measured in the upper panel on
  spatial scales $\lesssim{}1~\rm{}pc$, as explained in
  Sec.~\ref{sec:density-gradients}.  The \emph{abscissa} indicates the
  fraction of mass measured in the interferometer data which we
  consider to be the dense gas tracer. \emph{Horizontal lines} are
  drawn for some regions for which only mass--size data are available
  (i.e., Perseus, Ophiuchus, and Sgr~B2). \emph{Shaded regions}
  highlight the parameter space occupied by spheres with indicated
  singular power--law density profiles. \emph{Green lines and markers}
  present data from reference objects outside the CMZ. The \emph{gray
    shaded region} contains solutions inconsistent with a fundamental
  assumption made in our analysis, i.e., that column densities
  decrease with increasing effective radius. \emph{Error bars} in both
  panels indicate the change in properties in case the mass would
  change by a factor~2.\label{fig:mass-size-analysis}}
\end{figure}

All of the CMZ molecular clouds studied here have an unusually high
average gas density. This is illustrated in
Fig.~\ref{fig:mass-size-analysis}~(top), where we compile data on gas
reservoirs in the clouds as described in
Sec.~\ref{sec:results-continuum}. As discussed in that section, the
data are subject due to systematic uncertainties in assumptions about
dust properties and other parameters. It is conceivable that all
dust--based mass measurements are in error by factors $\sim{}2$, which
would mean that all data are systematically shifted up or down from
their true location. Relative errors between clouds or different
spatial scales are less likely, though.

If we assume that the Orion~A cloud and molecular clouds in the CMZ
have similar 3D--geometries, then densities of CMZ clouds exceed those
of Orion~A by an order of magnitude at radii $\sim{}5~\rm{}pc$. For
Sgr~B2 the excess becomes a factor $\sim{}10^2$ at $5~\rm{}pc$
radius. This is remarkable, given that Orion~A is one of the densest
clouds in the wider solar neighborhood.

Note that this shows that the mass and density of the well--known
cloud G0.253+0.016 is not exceptional in the CMZ. Instead we find that
all major CMZ molecular clouds have very high masses and average
densities when explored at radii well above a parsec.

Specifically, in Fig.~\ref{fig:mass-size-analysis}~(top) we plot the
mass and size of the target clouds from the analysis of Herschel maps
explained in Sec.~\ref{sec:results-continuum} on spatial scales
$\ge{}0.7~\rm{}pc$. Specific numbers are taken from Table~1 of
\citetalias{kauffmann2016:gcms_i}. On smaller spatial scales we use the dendrogram analysis of
the combined SMA and APEX data. Here we first determine the smallest
radius of a resolved
fragment. Figure~\ref{fig:mass-size-analysis}~(top) then shows the
data for the most massive fragment that exceeds this smallest radius
by at most 10\%. The lines in Fig.~\ref{fig:mass-size-analysis}~(top)
effectively connect the mass--size measurement for the most massive
``core'' of $\sim{}0.1~\rm{}pc$ radius with the mass--size data of the
$\sim{}1~\rm{}pc$ radius ``clump'' and the cloud in which the core is
embedded.

On small spatial scales $\sim{}0.1~\rm{}pc$ we also include data for
Sgr~B2 extracted from a column density map presented by
\citet{schmiedeke2016:sgrB2-dust}. That study uses a range of
continuum emission data sets obtained using LABOCA, the SMA, Herschel,
and the VLA to obtain a three--dimensional model of the density
distribution via radiative transfer modeling. This in particular takes
the internal heating by star formation into account in a detailed
fashion.  Here we explore a map of the mass distribution as collapsed
along the line of sight.  The data point near 0.1~pc effective radius
in Fig.~\ref{fig:mass-size-analysis}~(top) is obtained by running a
dendrogram analysis on that map in the same fashion as done for the
SMA data presented in this paper.

For reference in Fig.~\ref{fig:mass-size-analysis}~(top) we illustrate
the conditions in Orion via the analysis of Herschel and Bolocam data
explained in Sec.~\ref{sec:results-continuum}. We also indicate the
mass--size threshold for high--mass SF from \citet{jenthu2010:irdcs},
$m(r)\ge{}870\,M_{\sun}\cdot{}(r/{\rm{}pc})^{1.33}$. Assuming a
spherical geometry, uniform density, and a mean molecular weight per
$\rm{}H_2$ molecule of 2.8 proton masses
\citep{kauffmann2008:mambo-spitzer}, the mean $\rm{}H_2$ particle
density is
\begin{equation}
n({\rm{}H_2}) = 3.5\times{}10^4 ~ {\rm{}cm^{-3}} \cdot
  \left( \frac{M}{10^4 \, M_{\sun}} \right) \cdot
  \left( \frac{r}{\rm{}pc} \right)^{-3} \, .
  \label{eq:mean-density}
\end{equation}
The Herschel--based masses and sizes on the largest scales of the CMZ
clouds (Table~1 of \citetalias{kauffmann2016:gcms_i}) yield average densities in the range
$(0.9~{\rm{}to}~1.8)\times{}10^4~\rm{}cm^{-3}$. These densities are
relatively high for Milky Way molecular clouds: Orion~A, for example,
has an average density of only $2.3\times{}10^3~\rm{}cm^{-3}$ for a
radius of 4.4~pc.

For reference we also indicate the properties of the Arches cluster,
one of the most massive stellar aggregates in the
CMZ. \citet{espinoza2009:arches} find a mass of
$2\times{}10^4\,M_{\sun}$ in an aperture of 0.4~pc radius. Note that
Sgr~B2 is the only CMZ region massive enough to form a cluster of this
density structure by simply converting a fraction $<1$ of the mass
residing at a fixed radius into stars. \citet{walker2015:dust-ridge}
analyze this point in more detail. They conclude that it is indeed
hard to form an Arches--like cluster in a single epoch of star
formation. \citeauthor{walker2015:dust-ridge} therefore suggest that
massive clusters might form in a number of successive star formation
events.

\section{Dense Gas on Small Spatial Scales\label{sec:dense-gas}}
\subsection{Unusually Shallow Density Gradients in CMZ
  Clouds\label{sec:density-gradients}}
One of the most puzzling aspects of CMZ clouds is that the embedded
dense cores of about 0.1~pc radius have relatively low densities,
compared to the high average densities inferred above
(Fig.~\ref{fig:mass-size-analysis}~[top]). The most massive cores in
CMZ clouds studied here have masses similar to, and sometimes below,
the one of the most massive core in Orion~A, i.e., the Orion~KL
region. A representative mass of $400\,M_{\sun}$ within 0.1~pc radius
gives a density of $1.3\times{}10^6~\rm{}cm^{-3}$. When compared to
the average density on large spatial scales, the density in Orion~A on
0.1~pc scale increases by a factor $\sim{}600$. In the CMZ clouds the
density increase is lower by about an order of magnitude. This means
that, compared to clouds like Orion, CMZ clouds have much more shallow
density gradients. In \citet{kauffmann2013:g0.253} we demonstrated
this trend for G0.253+0.016. The new data show that this is a general
trend for CMZ clouds.

The mass--size data in Fig.~\ref{fig:mass-size-analysis}~(top) can be
used to perform this analysis in more detail. We use the most massive
features at 0.1~pc (interferometer data) and 0.7~pc radius (peak mass
per Herschel beam) to measure mass--size slopes. These are defined as
$b={\rm{}d}\,\ln(m)/{\rm{}d}\,\ln(r)$, which becomes
$b\approx{}\ln(m[0.7~{\rm{}pc}]/m[0.1~{\rm{}pc}])/\ln(0.7~{\rm{}pc}/0.1~{\rm{}pc})$
in our situation. The resulting slopes are shown as the ordinate of
Fig.~\ref{fig:mass-size-analysis}~(bottom). For reference we also
measure the mass--size slope using the data on Orion~A shown in
Fig.~\ref{fig:mass-size-analysis}~(top). We further include slope
measurements for the Perseus and Ophiuchus molecular clouds in the
solar neighborhood ($d\le{}300~\rm{}pc$) from
\citet{kauffmann2010:mass-size-ii}. These clouds only form low--mass
stars. We find that, compared to other regions in the Milky Way, many
CMZ clouds have unusually steep mass--size slopes.

These steep mass--size slopes imply unusually shallow density
gradients. In \citet{kauffmann2010:mass-size-ii} we demonstrate that
singular spherical power--law density profiles
$\varrho{}\propto{}r^{-k}$ imply mass--size relations
$m\propto{}r_{\rm{}eff}^{3-k}$ if $k\le{}3$. A detailed analysis shows
that these trends are generally preserved in a variety of density
profiles, even if spherical symmetry does not apply. We can therefore
use the relation $k\approx{}3-b$ to develop an idea of the density
profiles found in CMZ clouds. Two such relations are indicated in
Fig.~\ref{fig:mass-size-analysis}~(bottom). We see that many CMZ
molecular clouds have relatively shallow density profiles compared to
other clouds in the Milky Way.

We find that Sgr~B2, Sgr~C, and Sgr~D have a structure that
is consistent with a density profile $\varrho{}\propto{}r^{-2}$, but
that all other clouds have much more shallow density laws. This sets
some CMZ clouds in a fascinating way apart from other Milky Way
clouds, but this finding also means that there is a diversity in CMZ
density gradients. Section~\ref{sec:cloud-evolution} interprets these
trends in the context of evolutionary differences among CMZ clouds.

We caution that there is a small chance that strong temperature
gradients inside the clouds affect our measurements of $b$. As an
example we consider the case that the temperature strongly decreases
with decreasing $r$, so that the mass at $r=0.1~\rm{}pc$ is
underestimated by a factor 2. In this case the true value of $b$ would
be smaller by a moderate factor
$\ln(2)/\ln(0.7~{\rm{}pc}/0.1~{\rm{}pc})\approx{}0.36$. The opposite
would hold for temperatures increasing with decreasing $r$, i.e., true
value of $b$ is larger than estimated here. Such trends could indeed
in principle affect the data shown in
Fig.~\ref{fig:mass-size-analysis}~(bottom) and explain why the sample
splits into populations with low and high $b$. Recall, however, that
many CMZ dust emission peaks host star formation as evidenced by maser
emission (e.g., Fig.~1 of \citetalias{kauffmann2016:gcms_i}). It seems unlikely that the true
dust temperature drops significantly below the 20~K assumed here. In
this case we would not expect that the true value of $b$ deviates
significantly from the value reported in
Fig.~\ref{fig:mass-size-analysis}~(bottom). Also, since all CMZ clouds
host some star formation, it is likely that all CMZ clouds are
affected by similar errors in $b$. Temperature gradients alone are
then not likely to explain why CMZ clouds differ in $b$.

\subsection{Small Dense Gas Fractions in CMZ
  Clouds\label{sec:dense-gas-fraction}}
The aforementioned mass--size analysis characterizes the density
gradients of CMZ clouds. One other aspect of cloud structure is the
fraction of mass concentrated in the most massive and compact
features. This dense gas fraction is characterized here. We find
this property to be unusually small in many CMZ clouds. 

\emph{We stress that here we can only present approximate measures of
  the dense gas fraction.} The problem is that there are no clear--cut
definitions of total cloud mass and the mass of high--density gas. For
example, here we adopt the working definition that gas all material
detected by the interferometer resides in dense structures. This is
not necessarily true, and it is well possible that a low--density
filament or sheet seen edge--on will be just as detectable as denser
spherical cloud fragment. Here we derive a ratio between integrated
intensities instead of masses. The mass of dense gas is characterized
by integrating the flux in the \emph{interferometer--only} SMA images
that are \emph{not corrected for the primary beam pattern}. In this
calculation we only include emission exceeding the noise in the images
by a factor 2. We need to obtain a measure of the total cloud mass
that relates to the area imaged by the interferometer. Our measure is
based on the aforementioned ATLASGAL data
\citep{schuller2009:atlasgal} that are scaled to the observing
frequency of the SMA (see Appendix~A.1 of \citetalias{kauffmann2016:gcms_i}). Towards every CMZ
cloud we first multiply the ATLASGAL data with the sensitivity pattern
of the SMA interferometer observations. We then integrate the scaled
intensities to obtain a measure of the total cloud mass. The dense gas
fraction is then approximated by dividing the interferometer--only
flux density by the one derived from the scaled ATLASGAL images. This
ratio would be 1 if the compact structures picked up by the SMA
contained all the flux seen in the ATLASGAL maps.  This ratio is shown
on the abscissa of Fig.~\ref{fig:mass-size-analysis}~(bottom). We
obtain a reference value for the Orion~A molecular cloud using our
Bolocam map: the bolometer observations of a cloud at about 420~pc
distance very roughly approximate the spatial filtering induced by the
interferometer at 8.3~kpc distance. We stress that a more detailed
analysis is highly desirable but not straightforward. Specifically we
divide the mass obtained from our Bolocam map (assuming a temperature
of 15~K) by the Herschel--based total mass in
Fig.~\ref{fig:mass-size-analysis}~(top).

Figure~\ref{fig:mass-size-analysis}~(bottom) demonstrates that the
dense gas fraction varies within the CMZ, but that most CMZ clouds
have dense gas fractions that are by factors of 2~to~10 below the
reference value estimated for Orion~A. Similar conclusions were
reached by \citet{johnston2014:g0.253} and
\citet{rathborne2014:g0253-pdf}, who studied probability density
functions (PDFs) of the cloud column density\footnote{We abstain from
  obtaining column density PDFs from our data. The first reason is
  that it is very hard to gauge the impact interferometer--induced
  spatial filtering has on the data. Methods do exist to add extended
  emission back into the interferometer observations, but it is
  prudent to remember that some of these methods --- such as the
  FEATHER algorithm in CASA --- are approximate. These methods need to
  be tested against synthetic data before they can be trusted. The
  second reason is that the shape of column density PDFs depends on
  the cloud boundaries chosen for the study
  \citep{stanchev2015:column-density-pdf}. It is difficult and
  uncommon to factor this complication into the analysis of column
  density PDFs. This is a particular problem for interferometer
  observations where material is selected and weighted by the
  interferometer's primary beam.}: some CMZ molecular clouds are
relatively free of dense gas that could form stars. Interestingly, the
Sgr~C and Sgr~D clouds, that stand out with mass--size slopes
unusually flat for CMZ clouds, also stand out in relatively high dense
gas fractions. We return to this point in
Sec.~\ref{sec:cloud-evolution} on evolutionary differences between CMZ
clouds.

\begin{figure}
\includegraphics[width=\linewidth]{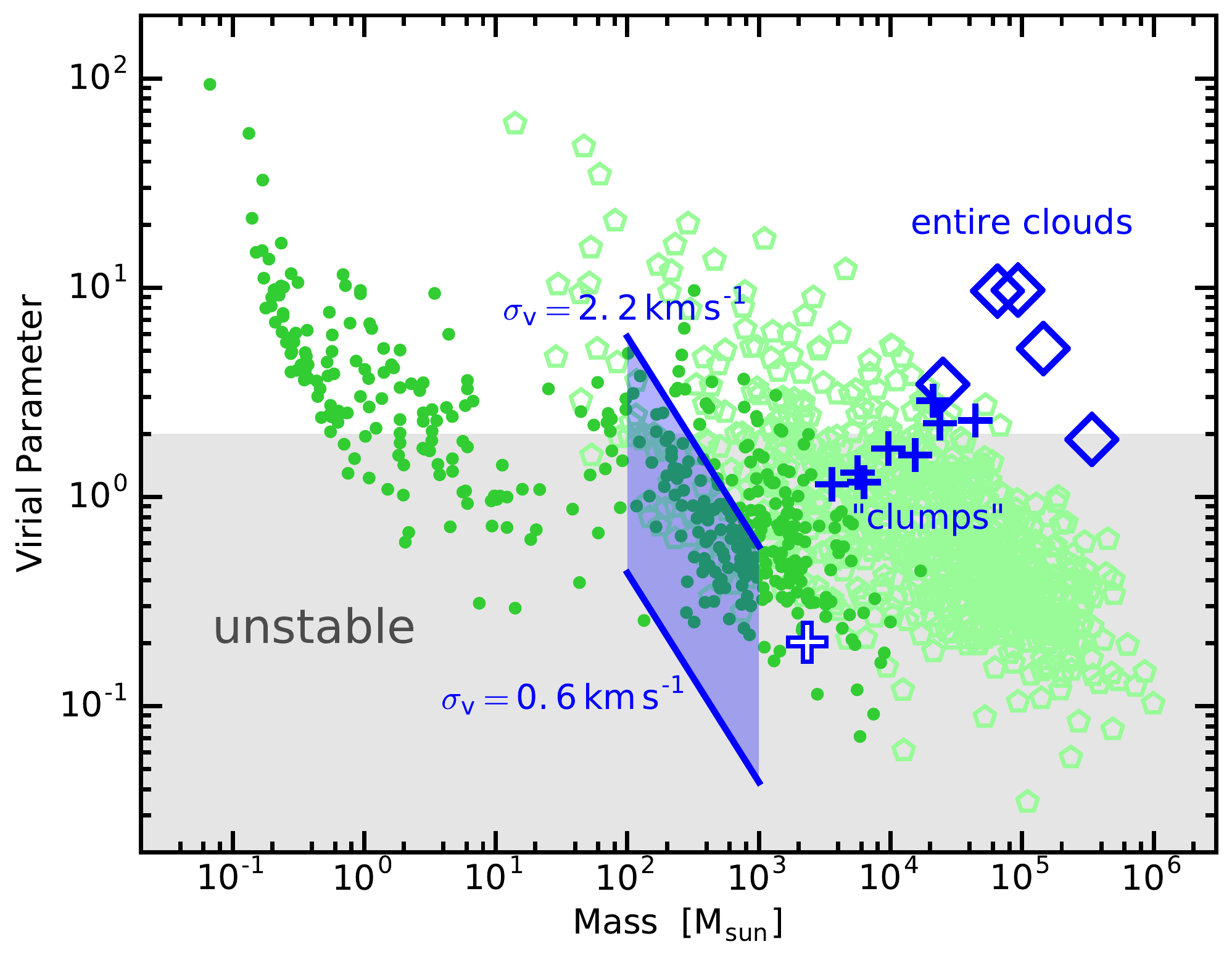}
\caption{Summary of the virial parameter measurements $\alpha$ from
  Sec.~\ref{sec:virial-parameter}. \emph{Blue symbols} give data for
  structures of varying spatial size explored in this paper. See
  Sec.~\ref{sec:line-emission} for terminology and details. The
  \emph{blue cross with white filling} indicates data for the Sgr~D
  region that probably resides outside the CMZ as studied here.
  \emph{Dark green bullets} indicate the properties of reference Milky
  Way clouds that were explored in emission lines of $\rm{}N_2H^+$ and
  $\rm{}NH_3$ tracing dense gas. Values for these are taken from the
  \citet{kauffmann2013:virial-parameter} compilation. \emph{Light
    green diamonds} give CO--based data for the lower density gas in
  Milky Way clouds reported in the same compilation. The \emph{blue
    shaded area} gives the virial parameter that would hold for cloud
  fragments of radius 0.1~pc that have masses and velocity dispersions
  within the ranges indicated in the figure. \emph{Gray shading} shows
  where $\alpha<\alpha_{\rm{}cr}\approx{}2$, indicating the domain
  where non--magnetized hydrostatic equilibria become unstable to
  collapse.\label{fig:virial-parameter}}
\end{figure}

\section{Gravitational Binding\label{sec:virial-parameter}}
In \citet{kauffmann2013:g0.253} we reasoned that the cloud
G0.253+0.016 is only marginally gravitationally bound on large spatial
scales, but that some of the embedded substructures might well be
bound and unstable to collapse. Here we confirm this finding and
establish it as a general trend for CMZ clouds (e.g.,
Fig.~\ref{fig:virial-parameter}).

The combined information on gas densities and kinematics allows to
calculate the virial parameter, $\alpha=5\,\sigma^2(v)\,R/(G\,M)$, as
defined by \citet{bertoldi1992:pr_conf_cores}. Measures of mass and
velocity dispersion refer to a given aperture of radius $R$. Clouds
are unstable to collapse if $\alpha<\alpha_{\rm{}cr}$. The critical
value depends on the nature of the pressure supporting the cloud, with
values $\alpha_{\rm{}cr}\approx{}2$ being appropriate in most cases
where magnetic fields are absent
\citep{kauffmann2013:virial-parameter}.

We caution that the concept of the virial parameter ignores several
processes that might be relevant in the CMZ. For example, the
gravitational potential of the CMZ can subject clouds to shear. The
fast internal motions of some clouds might already reflect this shear,
while in other clouds shear might just be about to set in and has not
yet increased internal motions. These shear motions will stabilize
clouds against collapse, and (depending on whether shear has set in or
not) the observed virial parameter might thus \emph{underestimate} a
cloud's ability to resist self--gravity. But CMZ clouds can also be
subject to \emph{compressive} tides (Sec.~\ref{sec:introduction}) that
increase self gravity. From this perspective the virial parameter
might \emph{overestimate} how well a cloud can withstand collapse. The
impact of shear and tides should be most significant on the largest
spatial scales. This suggests that virial parameter assessments on
entire clouds must be treaded with caution, while data on smaller
structures can be interpreted more reliably.

Figure~\ref{fig:virial-parameter} presents a virial analysis that uses
the information on cloud kinematics from \citetalias{kauffmann2016:gcms_i} (see
Sec.~\ref{sec:line-emission}). Given the range of velocity dispersions
(Fig.~7 of \citetalias{kauffmann2016:gcms_i}) and mass reservoirs
(Fig.~\ref{fig:mass-size-analysis}~[top]) on small spatial scales, we
adopt ranges for velocity dispersions and masses instead of fixed
numbers. A radius of 0.1~pc is assumed for the smallest structures,
corresponding to the spatial scale on which the smallest
interferometer--detected structures are extracted in
Fig.~\ref{fig:mass-size-analysis}~(top). Note that no direct mass
measurements are available for the ``clumps'' of about 1~pc radius:
here we adopt a mean $\rm{}H_2$ column density of
$2\times{}10^{23}~\rm{}cm^{-2}$ to calculate a mass on the basis of
the measured radii. This is a plausible value, as can be gleaned from
Fig.~\ref{fig:mass-size-analysis}~(top). In this respect the
properties for the ``clumps'' in Fig.~\ref{fig:virial-parameter}
should be taken as an estimate instead of a real
measurement. \citet{walker2015:dust-ridge} obtain slightly lower
virial parameters for some of our clouds. Their results resemble those
we obtain for the ``clumps'' of intermediate size. It is plausible
that this is a consequence of their scheme to reject unrelated
velocity components. The method 
\citeauthor{walker2015:dust-ridge} use to measure velocity
dispersions resembles the one we use for our ``clumps''.

It appears that the clouds on largest spatial scales are unbound or
only marginally bound. It is thus possible that several of these
clouds will disperse in the future. The situation is markedly
different for the ``clumps'' with radii of order 1~pc. It seems
plausible or even likely that many of these structures will remain
bound. These are excellent conditions for current or future star
formation. The densest interferometer--detected cores are very likely
subject to signifiant gravitational binding, depending on how exactly
gas motions and density structure
combine. \citet{rathborne2015:g0253-alma} and \citet{lu2015:20kms}
obtain similar results for G0.253+0.016 and the $20~\rm{}km\,s^{-1}$
cloud, respectively.

In summary the data presented here suggest that the conditions are
conducive for star formation on small spatial scales, but that it is
not clear that the entire clouds will participate in this
process. Note, however, that this analysis ignores magnetic fields (as
well as tidal forces and shear). \citet{pillai2015:magnetic-fields}
show that, at least in G0.253+0.016, magnetic forces due to a field
$\sim{}5.4~\rm{}mG$ are an important and possibly dominating factor in
the total energy budget. This might also hold for other CMZ
clouds. Magnetic fields have the effect of reducing the critical
virial parameter \citep{kauffmann2013:virial-parameter}. Provided
magnetic fields have a sufficient strength, this might mean that the
physical conditions in CMZ clouds are not favorable for star
formation.

\section{Discussion: Suppressing CMZ Star
  Formation\label{sec:cloud-evolution}}
We return to the problem illustrated in Fig.~\ref{fig:sf-mass}, i.e.,
the question why CMZ star formation in dense gas is suppressed by a
factor $\gtrsim{}10$ compared to other regions in the Milky Way. Here
we collect some of the factors examined above and interpret them in
context.

\subsection{Moderate CMZ Star Formation Density
  Thresholds\label{sec:sf-threshold}}
The most straightforward explanation of suppressed CMZ star formation
would be the existence of a high density threshold for star formation
that is not overcome by CMZ
clouds. \citet{kruijssen2013:sf-suppression-cmz} explore this option
on the basis of the observations by \citet{longmore2012:sfr-cmz}.
They conclude that an SF threshold $\rm{}H_2$ particle density
$\gtrsim{}10^7~\rm{}cm^{-3}$ is needed to explain the observed level
of SF suppression, provided steep power--law probability density
functions (PDFs) of gas density prevail in clouds. Log--normal
distributions, which are more commonly observed in molecular clouds
(e.g., \citealt{kainulainen2009:column-density-pdf}), imply threshold
densities
$\gtrsim{}3\times{}10^8~\rm{}cm^{-3}$. \citeauthor{kruijssen2013:sf-suppression-cmz}
also argue that the fast supersonic motions in CMZ clouds would in
fact imply threshold densities $\approx{}3\times{}10^7~\rm{}cm^{-3}$,
based on single--dish observations of clouds. Here we re--examine this
latter claim based on our independent measurements of gas density and
line width. In particular, we include the observations of low velocity
dispersions in our interferometer maps.

\citet{kruijssen2013:sf-suppression-cmz} build their analysis on
previous work by \citet{krumholz2005:general-sf-theory} and
\citet{padoan2011:sf-supersonic-mhd}. These authors argue that the
threshold $\rm{}H_2$ particle density for star formation in a cloud of
one--dimensional Mach number $\mathcal{M}$ exceeds the mean cloud
density by a factor
\begin{equation}
  \frac{n_{\rm{}SF,lim}}{\langle{}n\rangle} = (1.2\pm{}0.4) \cdot
  \alpha \, \mathcal{M}^2 \, ,
  \label{eq:sf-threshold}
\end{equation}
where $\alpha$ is the virial parameter. The range in the
proportionality constant reflects the differences between the
theoretical derivations by different groups. We adopt an approximate
value of 1.2 in the
following. \citet{kruijssen2013:sf-suppression-cmz} use average
properties of CMZ clouds gleaned from single--dish data to evaluate
this relation. They adopt Mach numbers of order 30, average densities
of a few $10^4~\rm{}cm^{-3}$, and virial parameters of order 1. This
yields the aforementioned values
$n_{\rm{}SF,lim}\approx{}3\times{}10^7~\rm{}cm^{-3}$.  The GCMS
provides more single--dish data on clouds and expands the relevant
observational picture because the interferometer observations reveal
rather narrow lines of $0.6~{\rm{}to}~2.2~\rm{}km\,s^{-1}$ on spatial
scales $\sim{}0.1~\rm{}pc$.

First consider entire clouds. In \citetalias{kauffmann2016:gcms_i} we show that
$\mathcal{M}=\sigma(v)/\sigma_{{\rm{}th},\langle{}m\rangle{}}(v)$ is
between 15 and 40 for our target clouds on spatial scales probed by
single--dish data, where $\sigma_{{\rm{}th},\langle{}m\rangle{}}(v)$
is the thermal velocity dispersion (here evaluated at 50~K) at the
mass of the mean free particle, $\langle{}m\rangle{}=2.33\,m_{\rm{}H}$
(see \citetalias{kauffmann2016:gcms_i}). We substitute the specific observed values of $\alpha$
and $\mathcal{M}$ for our target clouds in
Eq.~(\ref{eq:sf-threshold}). This gives
$n_{\rm{}SF,lim}=(0.1~{\rm{}to}~2)\times{}10^8~\rm{}cm^{-3}$. These
numbers are broadly consistent with the results by
\citet{kruijssen2013:sf-suppression-cmz}.

Alternatively we can use\footnote{ We remark that
  Eq.~(\ref{eq:sf-threshold}) was initially developed to describe the
  behavior of entire clouds, based on the properties prevailing on the
  largest spatial scales. However, it is plausible to assume that
  cloud fragments embedded in a larger complex follow the same
  relation. Virial parameter and Mach number do, after all, only set
  the boundary conditions for processes acting on much smaller spatial
  scales.} Eq.~(\ref{eq:sf-threshold}) to evaluate the SF threshold
density for structures of size $\sim{}0.1~\rm{}pc$. This is
particularly interesting because \citetalias{kauffmann2016:gcms_i} derives much lower velocity
dispersions and Mach numbers for these structures. To do this
evaluation we substitute the definitions of the thermal velocity
dispersion (Eq.~1 from \citetalias{kauffmann2016:gcms_i}) and the virial parameter in
Eq.~(\ref{eq:sf-threshold}). We obtain
\begin{equation}
  \frac{n_{\rm{}SF,lim}}{\langle{}n\rangle} \approx
  8.1 \,
  \left( \frac{\sigma(v)}{\rm{}km\,s^{-1}} \right)^4 \,
  \left( \frac{T_{\rm{}gas}}{50~\rm{}K} \right)^{-1} \,
  \left( \frac{R}{0.1~\rm{}pc} \right) \,
  \left( \frac{M}{100\,M_{\sun}} \right)^{-1} \, ,
\end{equation}
where $T_{\rm{}gas}$ is the gas temperature. We pick a representative
mass of $400\,M_{\sun}$ for the mass of structures of 0.1~pc
radius. Figure~\ref{fig:mass-size-analysis}~(top) shows that the
actual masses differ by a factor only $\sim{}2$ from this, if we
exclude the much more massive structures in Sgr~B2. Given the
uncertainty in velocity dispersions on this spatial scale, we
vary\footnote{Recall that we cannot obtain velocity dispersions for
  individual structures because there is no straightforward
  correspondence between dust and line emission (\citetalias{kauffmann2016:gcms_i}). This forces
  us to consider ranges in the parameters instead of specific
  observations for every cloud structure.}  $\sigma(v)$ between 0.6
and $2.2~\rm{}km\,s^{-1}$. For these parameters we find that
$n_{\rm{}SF,lim}/\langle{}n\rangle\approx{}0.3~{\rm{}to}~47$. A mean
density $\langle{}n\rangle=1.4\times{}10^6~\rm{}cm^{-3}$ holds for
0.1~pc radius and $400\,M_{\sun}$ mass
(Eq.~\ref{eq:mean-density}). This implies values of $n_{\rm{}SF,lim}$
between $4\times{}10^5~\rm{}cm^{-3}$ and $7\times{}10^7~\rm{}cm^{-3}$.
A representative velocity dispersion is arguably given by the median
value of the smallest velocity dispersions found in the clouds,
$\min(\sigma_{\rm{}int}[v])$ (Table~5 of \citetalias{kauffmann2016:gcms_i}). This median value
is $1.1~\rm{}km\,s^{-1}$. Choosing this velocity dispersion gives
$n_{\rm{}SF,lim}/\langle{}n\rangle=3$ and thus
$n_{\rm{}SF,lim}=4\times{}10^6~\rm{}cm^{-3}$. This latter result is at
the lower limit of the range in SF threshold densities considered by
\citet{kruijssen2013:sf-suppression-cmz}.

Note, however, that none of these considerations take the role of the
magnetic field into account. Section~\ref{sec:magnetic-field-sf}
discusses that this field might play a major role in shaping clouds.

In summary we generally derive SF threshold densities consistent with
what \citet{kruijssen2013:sf-suppression-cmz} estimate via
Eq.~(\ref{eq:sf-threshold}) from single--dish data. The interferometer
data may, however, hint at SF threshold densities that are as low as
$10^6~\rm{}cm^{-3}$, which is well below densities $10^7~\rm{}cm^{-3}$
deemed plausible before. Given our results, it appears plausible to
adopt $n_{\rm{}SF,lim}=10^{7~{\rm{}to}~8}~\rm{}cm^{-3}$ for the
CMZ. \citet{kruijssen2013:sf-suppression-cmz} estimate that
$n_{\rm{}SF,lim}\approx{}10^4~\rm{}cm^{-3}$ in the solar
neighborhood. The values for the CMZ are thus much higher, and this
difference might --- relative to the solar neighborhood --- help to
suppress CMZ star formation as observed. However, the estimated values
of $n_{\rm{}SF,lim}=10^{7~{\rm{}to}~8}~\rm{}cm^{-3}$ are at the lower
limit of the threshold densities between $10^7~\rm{}cm^{-3}$ and
$\gtrsim{}3\times{}10^8~\rm{}cm^{-3}$
\citet{kruijssen2013:sf-suppression-cmz} require to suppress CMZ star
formation at the observed level. For example, rather steep power--law
PDFs need to prevail to bring the predicted threshold densities from
Eq.~(\ref{eq:sf-threshold}) in agreement with the thresholds
\citeauthor{kruijssen2013:sf-suppression-cmz} estimate from the
observed SF. This suggests to look for additional factors that might
suppress CMZ star formation.

\begin{figure}
\includegraphics[width=\linewidth]{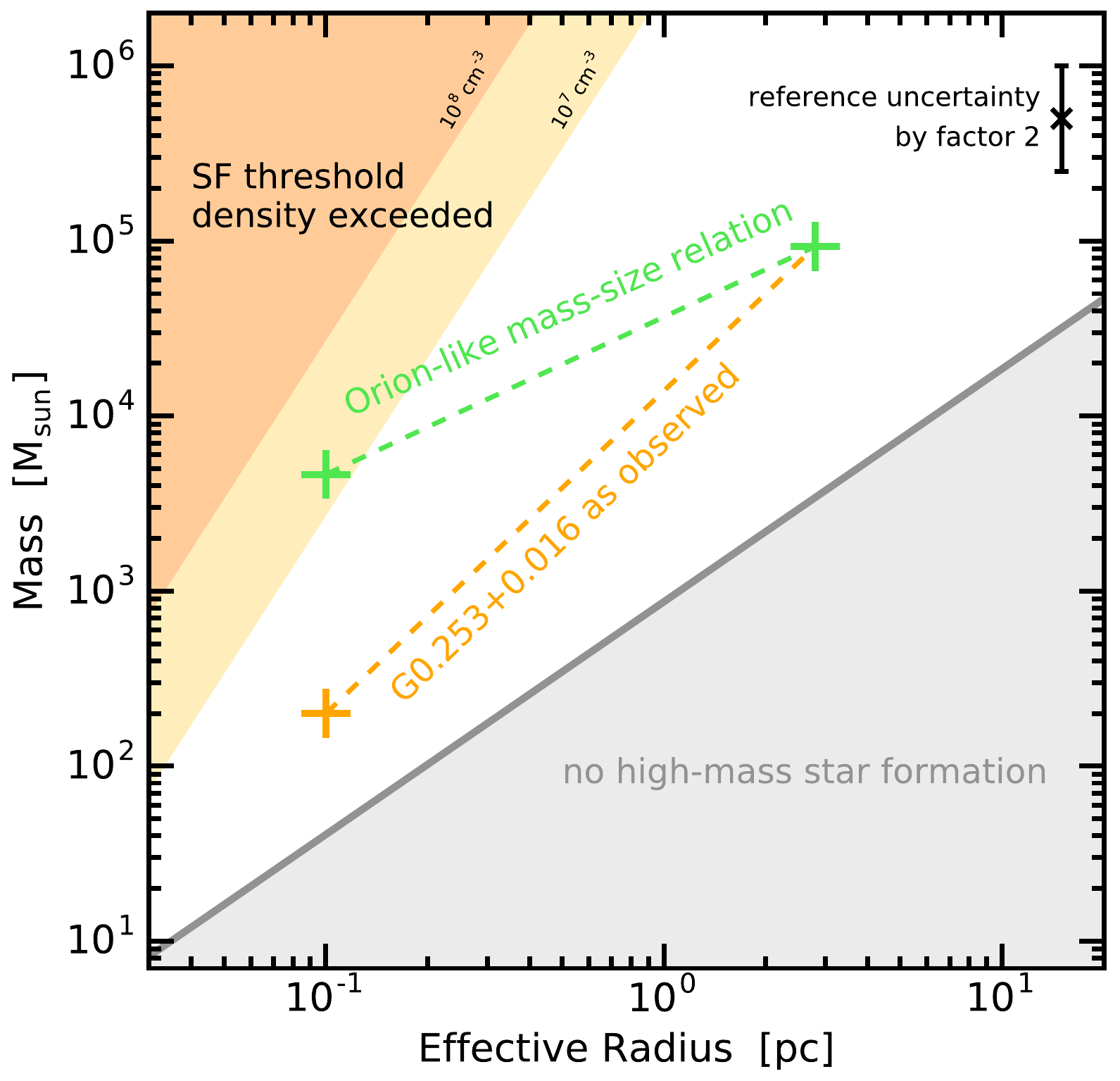}
\caption{The impact of mass--size slopes --- i.e., density gradients
  --- on CMZ star formation. The \emph{gray shading} is taken from
  Fig.~\ref{fig:mass-size-analysis}. The \emph{brown--orange shading}
  indicates where spheres would exceed the star formation density
  threshold. The shading gradient indicates the uncertainty in
  $n_{\rm{}SF,lim}$, here taken to be in the range
  $10^{7~{\rm{}to}~8}~\rm{}cm^{-3}$. The \emph{orange cross and dashed
    line} indicates the mass--size trend observed for
  G0.253+0.016. The \emph{green crosses and dashed line} gives the
  mass--size relation of a model cloud that has the same total mass
  and size as G0.253+0.016, but follows a more shallow mass--size
  relation (i.e., steeper density gradient), here taken from
  observations of Orion. The \emph{error bar} indicates the a
    representative uncertainty in mass measurements by a factor~2
    (Sec.~\ref{sec:densities}). In summary this diagram shows that
  shallow CMZ density laws help to suppress star formation because
  little or no gas at all exceeds
  $n_{\rm{}SF,lim}$.\label{fig:mass-size-context}}
\end{figure}

\subsection{Suppressed SF from Shallow Density
  Gradients\label{sec:mass-size-sf}}
It appears that the cloud density structure plays a key role in
determining the star formation rate of the CMZ clouds. Specifically,
it emerges that the shallow density gradients --- and corresponding
steep mass--size relations --- found in
Sec.~\ref{sec:density-gradients} are essential to understand CMZ star
formation. This is illustrated in Fig.~\ref{fig:mass-size-context}.

Section~\ref{sec:sf-threshold} yields SF threshold densities
$n_{\rm{}SF,lim}=10^{7~{\rm{}to}~8}~\rm{}cm^{-3}$. This range is
indicated in Fig.~\ref{fig:mass-size-context}. Now consider
G0.253+0.016 as an example. The observed mass--size trend indicated in
Fig.~\ref{fig:mass-size-context} steers clear of the domain where the
SF threshold density would be exceeded. For comparison
Fig.~\ref{fig:mass-size-context} also shows the mass--size trend
G0.253+0.016 would have if it had the same mass--size slope as Orion~A
(that value is taken from Fig.~\ref{fig:mass-size-analysis}). Here we
see that the density in G0.253+0.016 would then actually reach values
close to the SF density threshold.  Similar trends hold for all other
CMZ clouds explored here. This excludes the notable exception of
Sgr~B2, which reaches densities well in excess of the thresholds
considered here.

For this reason we argue that \emph{the low star formation activity of
  CMZ clouds is chiefly a consequence of the unusually shallow density
  gradients of CMZ clouds}. Theoretical research into CMZ cloud
structure must begin to explore this critical trend.

We speculate that shallow density gradients can emerge when the clouds
are not tightly bound by self--gravity. In that case gravitational
forces cannot help in concentrating mass into compact structures. This
has the consequence that cloud fragments much smaller than the total
cloud size will only contain a relatively small fraction of the
total cloud mass. In other words: such a region has a relatively steep
mass--size relation, implying shallow density gradients.
Figure~\ref{fig:virial-parameter} shows indeed that CMZ clouds are at
best marginally bound by self--gravity. This supports the picture
outlined above. Further, in Sec.~\ref{sec:introduction} we describe
that CMZ clouds are subject to compressive tides and are further
confined by significant thermal pressure that forces any gas at
moderate temperatures $\lesssim{}100~\rm{}K$ to reside at densities
$\gtrsim{}10^4~\rm{}cm^{-3}$. These two mechanisms might help to
produce massive clouds of high average density that are unbound.

\subsection{Suppressed SF from Strong Magnetic
  Fields\label{sec:magnetic-field-sf}}
The strong magnetic field penetrating CMZ clouds might be another
important factor in the suppression of CMZ star formation. The gas is
pervaded by a strong magnetic field of a few $10^3~\rm{}\mu{}G$
\citep{yusef-zadeh1984:non-thermal,
  uchida1985:cmz-lobes,chuss2003:cmz-polarization,
  novak2003:cmz-polarization}. This field also penetrates individual
CMZ clouds with a strength that is high enough to prevent the global
collapse of clouds \citep{pillai2015:magnetic-fields}. It is
conceivable that the shallow density profiles described above are a
direct consequence of magnetic fields dominating over self--gravity as
well as turbulence.

The observational properties of the CMZ magnetic field are well
established at this point (see, e.g.,
\citealt{morris2014:magnetic-field} for a review). What is less clear
is the origin of this strong field, and how exactly it couples to the
orbit on which clouds travel around in the CMZ. This makes it hard to
gauge the exact role the magnetic field might play for the evolution
of CMZ molecular clouds.

\subsection{A Note of Caution: Disconnects between SF and Density
  Structure}
We note that the current cloud structure can have little relation to
the recent star formation activity. Consider the
$50~\rm{}\,km\,s^{-1}$ cloud. As evident from the dust continuum
emission maps in \citetalias{kauffmann2016:gcms_i}, and quantified by the mass--size data in
Fig.~\ref{fig:mass-size-analysis}, this cloud is devoid of evidence
for dense and massive clumps that are definitely needed to form
high--mass stars. At the same time the cloud hosts four
\ion{H}{ii}~regions with ages of about 1~Myr or less (Table~5 and
Appendix~C of \citetalias{kauffmann2016:gcms_i}). This finding suggests that the cloud density
structure has evolved significantly since it gave birth to the
high--mass stars powering the \ion{H}{ii}~regions. This is plausible:
cloud structure can evolve on a time scale similar to the flow
crossing times $\sim{}3\times{}10^5~\rm{}yr$ obtained in Sec.~3.4 of
\citetalias{kauffmann2016:gcms_i}, which is shorter than the aforementioned lifetimes of
\ion{H}{ii}~regions.

In other CMZ clouds we find a more direct connection between the young
stars and dense cloud fragments. For example, \citet{lu2015:20kms}
study the $20~\rm{}km\,s^{-1}$ cloud and detect a population of water
masers closely associated with dense structures seen in SMA maps of
dust emission.

\begin{figure*}
\centerline{\includegraphics[width=\linewidth]{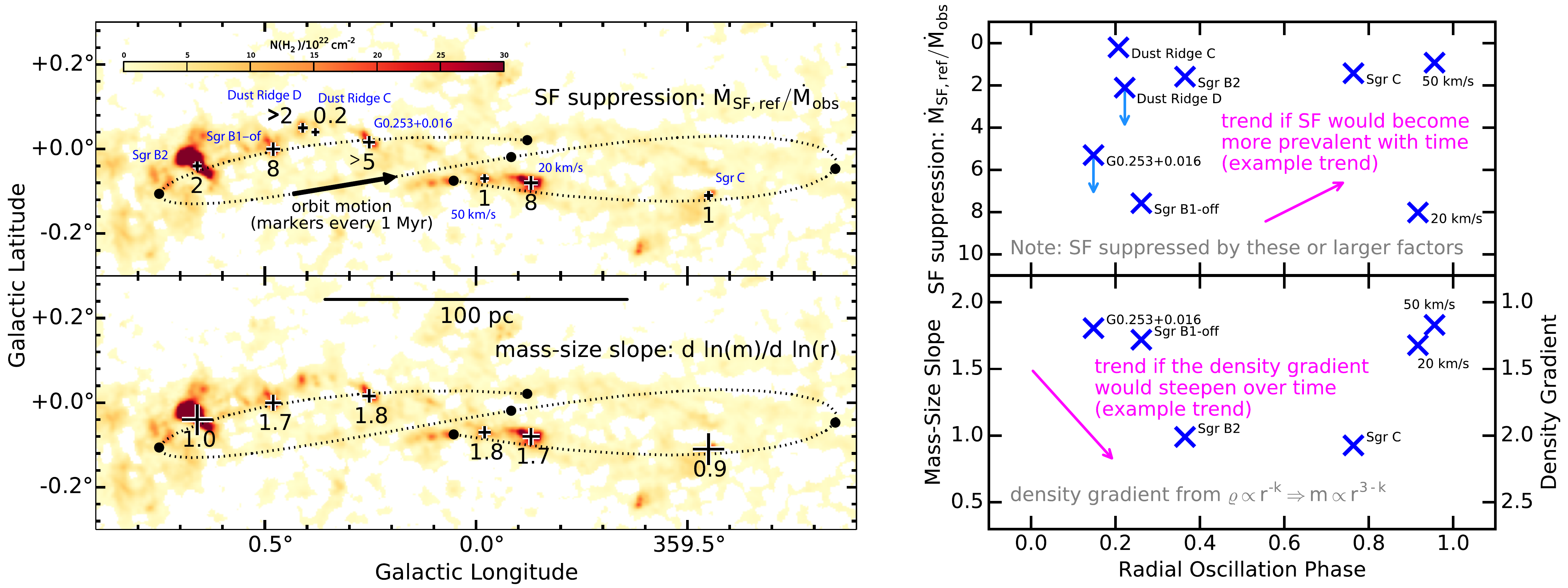}}
\caption{The spatial distribution of cloud properties. \emph{Blue
    labels in the top map} give cloud names. \emph{Symbol sizes and
    labels in the top map} indicate the factor by which the star
  formation activity in the dense gas is suppressed relative to a
  typical Milky Way reference value provided by
  Eq.~(\ref{eq:sf-reference}). \emph{Symbol sizes and labels in the
    bottom map} give the mass--size slope derived from the cloud
  structure data on spatial scales $\lesssim{}1~\rm{}pc$. A
  \emph{dotted line in the maps} shows part of the orbit for CMZ
  clouds proposed by \citet{kruijssen2014:orbit}. The \emph{background
    maps} present a column density map derived from Herschel dust
  emission data, as derived in this paper. The \emph{panels on
      the right} present the same information on SF suppression and
    mass--size slopes as a function of the radial phase of the orbit
    proposed by \citet{kruijssen2014:orbit}. The \emph{magenta arrows}
    indicate very roughly how a property might change, e.g., if SF
    suppression would decrease over time or in case density gradients
    would steepen over time: the directions of arrows matter (i.e., up
    or down vs.\ phase), but their position or placement does not. In
    summary the observed trends suggests that clouds do not follow a
    systematic evolutionary pattern as they orbit the
    CMZ.\label{fig:cloud-properties}}
\end{figure*}

\subsection{No systematic Cloud Evolution along CMZ
  Orbit\label{sec:trends-along-orbit}}
Figure~\ref{fig:cloud-properties} presents the mass--size slopes and
star formation rates in the context of the orbit of CMZ clouds
inferred by
\citet{kruijssen2014:orbit}. \citet{longmore2013:cmz-cluster-progenitors}
suggest that clouds evolve as they move along the orbit. In particular
the latter group proposes a sequence in which dense clouds form (or
get strongly compacted) near the location of G0.253+0.016 via
compression induced by the gravity of the CMZ potential centered on
the $\rm{}Sgr~A^{\ast}$ region, start to evolve towards star
formation, and become efficient star--forming clouds as they move
along the orbit towards Sgr~B2. In other words, the pericenter passage
sets an absolute time marker in these models of cloud evolution. Here
we cannot evaluate whether the pericenter passage affects the
structure of the clouds. However, our data only provide mixed evidence
for scenarios in which the pericenter passage induces a systematic
evolution towards SF with increasing orbital phase (measured with respect to
the pericenter location).

Figure~\ref{fig:cloud-properties} places our data into the overall
orbital context of the CMZ. We use
Fig.~\ref{fig:cloud-properties}~(left) to present the data in the
context of a map, while Fig.~\ref{fig:cloud-properties}~(right) shows
the same information but against the orbital phase. Specifically we
consider the orbital oscillations in radius, that have a period of
$2.03_{-0.18}^{+0.70}~\rm{}Myr$ (\citealt{kruijssen2014:orbit}; the
other periods are $3.69_{-0.30}^{+0.68}~\rm{}Myr$ in azimuth and
$2.27_{-0.34}^{+0.70}~\rm{}Myr$ perpendicular to the Galactic Plane,
respectively). The radial orbital phase is measured with respect to
the pericenter passage. The time since pericenter passage is taken
from Table~2 of \citet{kruijssen2014:orbit}.

Figure~\ref{fig:cloud-properties}~(top) summarizes the ratio
$\dot{M}_{\rm{}SF,ref}/\dot{M}_{\rm{}obs}$ between the observed star
formation rate, $\dot{M}_{\rm{}obs}$, and the reference value taken
from \citet{lada2010:sf-efficiency},
\begin{equation}
\dot{M}_{\rm{}SF,ref} = (4.6\pm{}2.6)\times{}10^{-8}\,M_{\sun}\,{\rm{}yr}^{-1}
  \cdot{}(M_{\rm{}dense}/M_{\odot}) \, .
\label{eq:sf-reference}
\end{equation}
\citeauthor{lada2010:sf-efficiency} take $M_{\rm{}dense}$ to be the
cloud mass residing at a column density corresponding to a visual
extinction $A_V\gtrsim{}7~\rm{}mag$. The ratio
$\dot{M}_{\rm{}SF,ref}/\dot{M}_{\rm{}obs}$ indicates the factor by
which star formation in a region is suppressed. Here we use the total
cloud masses from Table~1 in \citetalias{kauffmann2016:gcms_i},
representing masses at $A_V\gtrsim{}100~\rm{}mag$, to evaluate
Eq.~(\ref{eq:sf-reference}). The star formation rates are taken from
Table~7 in \citetalias{kauffmann2016:gcms_i}. These latter estimates
are largely based on the number of radio--detected \ion{H}{ii}~regions
that are embedded in a cloud.

We do find an increase in the dense gas star formation efficiency
between G0.253+0.016 and Sgr~B2 by a factor $\gtrsim{}3$. However,
current data do not indicate a monotonic increase of the star
formation activity, as indicated by the case of the Dust~Ridge~C cloud
where $\dot{M}_{\rm{}obs}$ is up to $5\,\dot{M}_{\rm{}SF,ref}$ and the
case of significantly suppressed star formation in the Sgr~B1--off
region between Sgr~B2 and Dust~Ridge~C.

The situation is even less clear --- though not necessarily in
conflict with the \citet{longmore2013:cmz-cluster-progenitors} model
--- when considering star formation in the $20~\rm{}km\,s^{-1}$ and
$50~\rm{}km\,s^{-1}$ clouds. First, the locations of these clouds
along the proposed orbit are separated by only about $10^5~\rm{}yr$,
while their SF rate per unit dense gas varies sharply by a factor
$\sim{}9$. It is difficult to imagine how this difference in SF rate
should emerge over such a small time scale. Second, the
$50~\rm{}km\,s^{-1}$ cloud show significant star formation
\emph{ahead} of the pericenter passage. That said,
\citet{longmore2013:cmz-cluster-progenitors} only speculate that their
model holds for clouds between G0.253+0.016 and Sgr~B2, and
\citet{kruijssen2014:orbit} show that the $20~\rm{}km\,s^{-1}$ and
$50~\rm{}km\,s^{-1}$ clouds are disconnected from the aforementioned
sequence of clouds. Still, the $20~\rm{}km\,s^{-1}$ and
$50~\rm{}km\,s^{-1}$ clouds highlight that the cloud state before
pericenter passage will influence the subsequent cloud evolution.

Figure~\ref{fig:cloud-properties}~(bottom) presents measures of the
gas density distribution, i.e., the mass--size slope. We do find
noteworthy variations within the CMZ: Sgr~B2 and Sgr~C have slopes
significantly different from all other clouds along the proposed orbit
studied here. However, there is only weak evidence for any systematic
trend. In particular we see no change between G0.253+0.016 and
Sgr~B1--off: such a trend would probably be expected if clouds were
evolving towards star formation between G0.253+0.016 and Sgr~B2. That
said, the Sgr~B2 region has the shallow mass--size slopes naively
expected for regions that evolve towards star formation by increasing
their density gradients. Still, if the difference between Sgr~B1--off
and Sgr~B2 is a result of evolution, then this process must be
completed in the about $3\times{}10^5~\rm{}yr$ that it takes to travel
between the clouds along the orbit. This would be an exceptionally
fast evolution in cloud structure, comparable to the dynamic crossing
times for CMZ clouds (Sec.~3.4 of \citetalias{kauffmann2016:gcms_i}).

Note that the  \emph{inverse} trend is observed for the evolution
between Sgr~C and the $50~\rm{}km\,s^{-1}$ cloud, i.e., the mass--size
slope increases along the orbit. This is clearly not expected for
straightforward evolution towards star formation. Some additional
hypotheses are required to explain this trend.

We stress that some of this discussion depends on whether all CMZ
clouds do indeed follow the orbit proposed by
\citet{kruijssen2014:orbit}. Their model provides a good mathematical
description of the structure of the CMZ. Note, however, that some
studies find evidence for interaction of the $20~\rm{}km\,s^{-1}$ and
$50~\rm{}km\,s^{-1}$ with the inner Galactic Center environment. See
\citet{herrnstein2005:inner-5pc} for a discussion of the evidence,
including material taken from earlier sources. Such interactions would
place these two clouds within about 10~pc from
$\rm{}Sgr~A^{\ast}$. This would be inconsistent with the
\citeauthor{kruijssen2014:orbit} orbital model, and such deviant
clouds should not be placed in
Fig.~\ref{fig:cloud-properties}~(right).\medskip

\noindent{}This leaves us with a mixed record on evidence for an
evolutionary sequence along the \citet{kruijssen2014:orbit} orbit that
is primarily controlled by the orbital phase --- i.e., the separation
in space or time from the closest pericenter passage along the CMZ
orbit.  The spatial distribution of CMZ star formation does not
support this idea, while there is limited evidence from the analysis
of cloud density structure.

We stress that these observations complement the ideas forwarded by
\citet{longmore2013:cmz-cluster-progenitors} and
\citet{kruijssen2014:orbit}. First, as noted before, we cannot test whether a given cloud
is modified during pericenter passage, as proposed by
\citet{longmore2013:cmz-cluster-progenitors}. Second, we only explore
whether the orbital phase is the \emph{primary} parameter controlling
SF. It is well possible that factors like initial density, etc., also
play a role and that clouds \emph{do} follow an evolutionary sequence
as they orbit the CMZ --- but on an individual time line that is not
necessarily similar to the one of neighboring CMZ clouds. A need to
include properties of individual clouds would reduce the potential
value of an absolute time line (i.e., with respect to the pericenter
passage) and make model tests very complex. In this sense our
observations point towards the idea of
\citeauthor{kruijssen2014:orbit} (\citeyear{kruijssen2014:orbit};
their Sec.~4.2) that the evolution along the orbit is --- if at all
present --- to be seen in a \emph{statistical} sense: there can be
variation between clouds at given orbital phase, depending on initial
conditions. More measurements than accessible here are needed to
explore this statistical view.

\section{Summary\label{sec:summary}}
We present the first comprehensive study of the density structure of
several molecular clouds in the Central Molecular Zone (CMZ) of the
Milky Way. This is made possible by using data from the Galactic
Center Molecular Cloud Survey (GCMS), the first systematic study
resolving all major molecular clouds in the CMZ at interferometer
angular resolution (\citealt{kauffmann2016:gcms_i}; hereafter
\citetalias{kauffmann2016:gcms_i}).

We combine the new characterization of GCMS dust emission data with
information on the cloud kinematics and the star formation (SF)
activity from \citetalias{kauffmann2016:gcms_i}. This leads us to a number of conclusions.
\begin{itemize}
\item All major CMZ molecular clouds have high average densities
  $\sim{}10^4~\rm{}cm^{-3}$ when explored on spatial scales
  $\gtrsim{}1~\rm{}pc$ (Sec.~\ref{sec:densities}). This is illustrated
  in Fig.~\ref{fig:mass-size-analysis}~(top). The well--known cloud
  G0.253+0.016 (a.k.a.\ the ``Brick'') is one of several massive
  and dense clouds in the CMZ.
\item Many CMZ molecular clouds have unusually shallow density
  gradients (Sec.~\ref{sec:density-gradients}). This is reflected in
  mass--size slopes that are much steeper than what is found in, e.g.,
  SF regions in the solar neighborhood
  (Fig.~\ref{fig:mass-size-analysis}). For example, the structure of
  many CMZ clouds is consistent with --- highly simplistic ---
  power--law density profiles resembling $\varrho{}\propto{}r^{-1.3}$
  (Fig.~\ref{fig:mass-size-analysis}~[bottom]). In addition, a
  relatively small fraction of the cloud mass is concentrated in such
  dense structures (Sec.~\ref{sec:dense-gas-fraction},
  Fig.~\ref{fig:mass-size-analysis}~[bottom]).
\item Random ``turbulent'' gas motions imply an SF density threshold
  in the range $10^{7~{\rm{}to}~8}~\rm{}cm^{-3}$
  (Sec.~\ref{sec:sf-threshold}). These high densities probably help to
  suppress SF in the CMZ. A further critical factor for SF suppression
  appear to be the unusually shallow density gradients observed in
  many CMZ clouds. The resulting steep mass--size laws imply that very
  little mass in the clouds will exceed the SF density threshold
  (Sec.~\ref{sec:mass-size-sf}, Fig.~\ref{fig:mass-size-context}). We
  speculate that these shallow density laws are a consequence of weak
  gravitational binding of the clouds.
\item We find mixed evidence that clouds systematically evolve towards
  SF with increasing orbital phase along the CMZ orbit (i.e., with
  respect to pericenter passage;
  Sec.~\ref{sec:trends-along-orbit}). These observations can help to
  refine ideas about CMZ cloud evolution forwarded by
  \citet{longmore2013:cmz-cluster-progenitors} and
  \citet{kruijssen2014:orbit}. Specifically, we find no clear trend
  in star formation activity per unit dense gas
  (Fig.~\ref{fig:cloud-properties}~[top]), while we do find a trend in
  density gradients (i.e., mass--size slopes;
  Fig.~\ref{fig:cloud-properties}~[bottom]). However, the
  interpretation of this latter trend as a consequence of cloud evolution
  along the orbit requires massive changes of cloud structure on time
  scales of order 0.3~Myr, which we consider to be questionable.
\end{itemize}

\begin{acknowledgements}
  We thank a thoughtful and helpful referee who provided a thorough
  review full of insights. We also greatly appreciate help from
  Diederik Kruijssen and Steve Longmore who provided detailed feedback
  on draft versions of the paper. These combined sets of comments
  helped to significantly improve the quality and readability of the
  paper before and during the refereeing process. This research made
  use of astrodendro, a Python package to compute dendrograms of
  astronomical data (\path{http://www.dendrograms.org/}). JK and TP
  thank Nissim Kanekar for initiating a generous invitation to the
  National Centre for Radio Astrophysics (NCRA) in Pune, India, where
  much of this paper was written. QZ acknowledges support of the
  SI~CGPS grant on Star Formation in the Central Molecular Zone of the
  Milky Way.This research was carried out in part at the Jet
  Propulsion Laboratory, which is operated for the National
  Aeronautics and Space Administration (NASA) by the California
  Institute of Technology. AEG acknowledges support from NASA grants
  NNX12AI55G, NNX10AD68G and FONDECYT grant 3150570.
\end{acknowledgements}

\bibliographystyle{aa}
\bibliography{/Users/jens/texinputs/mendeley/library}
%\bibliography{mendeley/library}

\end{document}